\documentclass[16pt,a4]{article}
\usepackage{epsfig}
\usepackage{setspace}
\title{Parameters Influencing the Optical Properties Of SnS Thin Films}

\author{Priyal Jain$^{a,b}$ and P.
Arun$^b$\footnote{email:arunp92@physics.du.ac.in, Telephone:091 011
29258401, Fax: 091 011 27666220} \\ \\
$^a$Department of Electronic Science,\\University of Delhi-South Campus,
Benito Juarez Marg,\\ Delhi 110 021, INDIA \\ \\ 
$^b$Material Science Research Lab,\\ S.G.T.B. Khalsa College,
University of Delhi,\\ Delhi 110 007, India\\
}

\begin{document}
\maketitle

\begin{abstract}
Tin Sulphide (SnS) thin films have been recognized as a potential 
candidate for solar cells. Many fabrication techniques have been used to
grow SnS thin films. The band-gap, ${\rm E_g}$ of SnS films as reported in
literature, were found to vary from 1.2-2.5~eV depending on the film 
fabrication technique. The present work reports the structural, compositional, 
morphological and optical characterization of SnS thin films fabricated by 
thermal evaporation at room temperature. Results show that for the given 
fabrication technique/ condition, the band-gap functionally depends on the 
lattice parameter and grain size. The well-defined variation allows for 
tailoring SnS film as per requirement.

\end{abstract}

\vfil \eject

\section{Introduction}
Tin Sulphide (SnS) is formed from elements belonging to the
IV-VI group of the periodic table \cite{amartya}. Like other sulphide
compounds (${\rm As_2S_3}$ and ${\rm Sb_2S_3}$ {\sl etc.}), SnS too 
attracted attention in the 1970 and 1980's for possible application as an 
optical storage medium \cite{radot, patil}. However, recent investigations 
are directed towards studying its potential as a photovolatic or solar cell 
essentially due to its high absorption coefficient, low cost and low toxicity 
\cite{zhi,noguchi}. A pertinent point is that both applications require SnS 
to be in its thin film state. Being a layered chalcogenide 
\cite{nikol, albanesi} of relatively low melting point \cite{crc}, SnS has 
been found to be a suitable candidate for thin film fabrication. Depending on 
film thickness, the color of SnS films varies from orange to brown when 
viewed in transmission. Thus, the last 30-40 years has seen substantial work 
in fabrication of SnS thin films 
using various methods like thermal evaporation \cite{nahass}, RF sputtering
\cite{hartman}, chemical vapor deposition \cite{nair}, electrodeposition
\cite{ghaza} and spray pyrolysis \cite{thanga}. The reported optical band-gap 
of SnS films lies between 1.1-2.1~eV \cite{gao, sohila} depending 
on the fabrication method used. 

While there are large number of reports on SnS, more systematic studies 
should be undertaken considering its potential commercial application as a 
solar cell. Especially to understand why the band-gap of SnS varies so much. 
With this intention, we have  carried out systematic studies on SnS thin films 
of varying thicknesses grown by thermal evaporation. The structural, 
morphological, compositional and optical properties of the films are reported 
here. The role of the intra-planar forces in the layer structured (as seen from 
the lattice parameters of unit cells within the nano-crystalline SnS grains) 
and the grain size on the band-gap of the films have been discussed.

\section{Experimental}
SnS thin films of varying thickness were grown on glass substrates maintained 
at room temperature by thermal evaporation at vacuum better than ${\rm \sim 
4\times10^{-5}}$~Torr. A Hind High Vac (12A4D), Bengaluru thermal evaporation 
coating 
unit was used for the thermal evaporation process. SnS powder of 99\% purity 
supplied by Himedia (Mumbai) was used as the starting material. The 
thicknesses of the as grown films were then measured by a Veeco Dektak Surface 
Profiler (150). The standard structural, morphological and optical 
characterization of the films were done using Bruker D8 Diffractometer 
(in the ${\rm \theta-2\theta}$ mode), Renishaw Invia Raman Microscope, 
Field-Emission Scanning Electron Microscopy (FE-SEM FEI-Quanta 200F) and 
Systronics UV-VIS Double beam Spectrophotometer (2202). Chemical composition
of the films were determined using Kratos Axis Ultra DLD's X-Ray 
Photoelectron Spectroscopy (XPS) with ${\rm Al\,k\alpha}$ target.

\section{Results and Discussion}
\subsection{Structural Studies}

The X-Ray diffractograms of very thin SnS films (thickness less than 150~nm) 
were unmarked with a hump between (${\rm 2\theta\approx}$) ${\rm 10-60^o}$. 
The absence of peaks suggests that the samples were not crystalline. This is 
not surprising since chalcogenides grown at room temperature are usually 
amorphous in nature. However, the hump indicates that there might be some 
short range order among the atoms. As the sample thickness increases (270~nm 
and above), a small diffraction peak emerges on the hump around ${\rm 2\theta 
\approx 31^o}$. Thus, it would appear that ordering increases with increasing 
film thickness. Fig~(1a) shows the X-Ray diffraction patterns of 
samples of different thicknesses. The peak positions matched with those 
reported in ASTM Card No. 83-1758 implying an orthorhombic unit cell 
structure with lattice parameters of 4.148, 11.48 and 4.177~\AA. The standard 
relationship \cite{Cullity} was used
to determine the lattice parameters using the peak positions 
\begin{eqnarray}
sin^2\theta_i=Ah_i^2+Bk_i^2+Cl_i^2\nonumber
\end{eqnarray}
The lattice parameters `b' and `c' were found to be ${\rm 11.39~\AA \pm
0.01~\AA}$ and ${\rm 4.117~\AA \pm 0.002~\AA}$ respectively. The lattice 
parameters were found to remain constant with increasing film thickness 
within experimental error (fig~1b). Fig~(1b) however, shows an initial 
decreasing trend of lattice parameter `a' with increasing film thickness. 
However, for film thicknesses greater than 600~nm, there is no tensile stress 
acting on the films and `a' remains constant. 

Fig~2 is a Transmission Electron Microscope (TEM) image 
of a 480~nm thick film. The image clearly shows the
layered nature of the SnS films. The interplanar distances was found to be
2.45\AA\, (${\rm \approx b/4}$). This would mean the films are oriented with
the (040) plane lying parallel to the substrate. The oriented nature of the
films is also seen in the XRD patterns with the (040) 
peak standing out as the most intense peak in agreement with observations 
reported \cite{yue, Feng}. The X-Ray diffraction peak's 
Full Width at Half Maxima (FWHM) gives the grain 
size using Scherrer formula \cite{Cullity}. Fig~(3) shows a linear
relationship between the average grain size and film thickness. 

\subsection{Raman Spectra and XPS Analysis}

The Raman spectra of SnS thin films were collected using Argon laser in back 
scattering mode. The laser power (15~mW), exposure time (30~sec) and beam area 
were maintained constant for all the samples. Since SnS is known to show 
changes induced by photo-thermal absorption of light \cite{radot, patil}, all 
the Raman spectra were collected using low power laser. The nano-crystalline 
samples showed three prominent peaks which were consistently present in all 
the samples (fig~4) in varying proportions around ${\rm \sim 170}$, ${\rm 
\sim 230}$ and ${\rm \sim 330~cm^{-1}}$. However, each sample showed slight 
displacement in peak position compared to those reported for single crystal 
Raman peaks \cite{nikol}. The consensus in literature is that the 
${\rm 330~cm^{-1}}$ peak indicates existence of ${\rm SnS_2}$ \cite{Lucovsky}. This 
may have resulted due to the growth technique, since starting material's 
Raman does not exhibit peak at ${\rm 330~cm^{-1}}$. The ${\rm SnS_2}$ 
contribution decreases with thickness as is evident from the decreasing 
${\rm 330~cm^{-1}}$ peak intensity and area with increasing film thickness.

This was also confirmed by the chemical composition 
analysis done on the films of different thicknesses using X-Ray Photo-electron 
Spectroscopy (XPS). Fig~5 compares the XPS peaks of sulphur and tin for 270 
and 600~nm thicknesses. The 600~nm thick sample had lone sulphur 2p and tin 
${\rm 3d_{5/2}}$ peaks that could not be deconvoluted. The number of sulphur
(${\rm n_S}$) and tin atoms (${\rm n_{Sn}}$) in bonding per ${\rm cm^3}$ area
can be evaluated using the area under the curve (${\rm \Delta}$) and 
cross-sectional values (${\rm \sigma}$) \cite{cross} in the standard 
formula \cite{sigben}
\begin{eqnarray}
{n_s \over n_{Sn}}=\left({\Delta_S \over \sigma_S} \right)
\left({\sigma_{Sn} \over \Delta_{Sn}}\right)\nonumber
\end{eqnarray}
The ratio of sulphur to tin in this sample was found to be ${\rm \approx 0.9}$, 
suggesting within experimental error that the 600~nm sample is of SnS. 
However, in comparison with this, the 270~nm sample's sulphur (2p) and 
tin (${\rm 3d_{5/2}}$) peaks can be deconvoluted into 
two peaks. Based on the Raman results, we associate the major peaks with SnS 
and the minor ones with ${\rm SnS_2}$. Using the above mentioned formula and 
area under the curves, the ratio of S/Sn from the major peaks give 1.1 while 
from the minor peaks ${\rm S/Sn \approx 1.9}$. Thus, thin samples (270~nm and 
480~nm) have both species of SnS and ${\rm SnS_2}$. From the ratio of two tin 
peaks, we may say only ${\rm 10-12\%}$ of the film is ${\rm SnS_2}$. 

The ${\rm 170~cm^{-1}}$ peak or the ${\rm B_{2g}}$ peak is associated with
interaction along the inter-layer `b' axis \cite{Chandrasekhar} while the 
${\rm 238~cm^{-1}}$ peak is the symmetric Sn-S 
bond’s stretching mode (${\rm A_g}$) \cite{sohila}. As the film thickness 
increases, the 
position of these two peaks shifts (Fig~6 shows the variation). Initially,
the peak position of ${A_g}$ was found to decrease with increasing film 
thickness, however, after 600~nm the peak position remained fixed ${\rm
\approx 226~cm^{-1}}$. The region of saturation corresponds to where samples 
have constant lattice parameters. Hence, we believe that the peak 
position of ${\rm A_g}$ is intimately related to the lattice size. A plot 
between ${\rm A_g}$ and `a' (fig~6a) confirms a linear relation between the 
two. An increase in wave-number (vibration frequency) marks the increase in
restoring force (bond strength). Hence, the results suggest that increasing 
tensile stress along `a' direction is accompanied with increased restoring 
force acting between the tin and sulphur atom.

Fig~6 exhibits the variation in ${\rm B_{2g}}$ peak position with film 
thickness. This shows a monotonous decrease with film thickness. 
Shift in Raman peak position may also be related to grain size 
\cite{hyunchul}-\cite{s11} and nature of defect on the grain boundaries 
\cite{kuninori}. Fig~(7b) plots ${\rm B_{2g}}$ peak position with respect to 
the grain size. We find that the plot conforms to results of Zhu {\sl et al} 
\cite{zhu2} where a shift to lower wave numbers takes place with increasing 
grain size. As reported, SnS
are layered compounds with layers formed by `Sn' and `S' zig-zagged
molecules. These layers arrange themselves normal to the `b' axis with weak
Van der-Waal ``inter-planar" forces acting between them. The layers
themselves, due to this zig-zagging, have finite thickness with strong
``intra-planar" forces acting within it. If any stress acts along the `b'
axis, the ``inter-planar" distances vary with no effect on the
``intra-planar" distances \cite{Ehm}. Variation in grain size seems to
effect ``inter-planar" distances, thus effecting ${\rm B_{2g}}$ peak position.
Increased grain size is accompanied with lower Raman wave-numbers and hence
indicate lower restoring forces, which from our above analysis on ${\rm
A_g}$, would imply lattice constant `b' would decrease with increasing
grain size. However, the variation in `b' (${\rm 11.41-11.37\pm 0.01\AA}$) as 
seen from X-Ray Diffraction results) is too small to justify this comment.

\subsection{Morphological Studies}
Fig~(8) shows representative SEM images of SnS as grown films of different
thicknesses, namely 150, 480, 600 and 900~nm. The grain morphologies are
similar to those reported earlier \cite{r3}. In the case of the thicker
films (480, 600 and 900~nm), the morphology of the films appear very similar
to each other, with noticeable increase in grain density with 
increasing film thickness. However, the morphology of the 
150~nm thick film is distinctly different consisting spherical grains. As 
discussed above, Raman spectra (fig~4) and XPS analysis clearly shows 
that this sample contains ${\rm SnS_2}$. This should 
explain the distinctly different 
morphology of this sample from the remaining samples. Though results related
to this sample have not been used in this present study, we have included
its SEM to show very thin films are dominated with ${\rm SnS_2}$ marked by
spherical grains. Grains of these morphology are absent in samples with
thicknesses greater than 250~nm, highlighting, no or negligible contribution
by ${\rm SnS_2}$ in these samples.

\subsection{Optical Studies}

The absorption spectra of as grown SnS films with varying thicknesses were 
obtained between 300 and 1100~nm at room temperature. The band-gaps were 
evaluated using the absorption coefficient `${\rm \alpha}$' obtained from 
the UV-visible spectra using standard formula \cite{beer}
\begin{eqnarray}
\alpha =2.303 \left({A \over t}\right)\nonumber
\end{eqnarray}
where `A' and `t' are the film's absorbance and thickness respectively. The
band-gap of the films can be evaluated from this information. SnS is
reported to have both indirect \cite{sohila} and direct band-gap \cite{Devika}.
Band-gaps are usually evaluated using Tauc's method \cite{tauc} where a graph
between ${\rm (\alpha h\nu)^n}$ is plot with respect to ${\rm h\nu}$. The 
linear region of this plot is extrapolated to intersect the `X'-axis at y=0. 
The point of intersection gives the band-gap of the material. For 
direct allowed transitions, the absorption coefficient is related to the 
photon energy by equation \cite{mott}-\cite{clark}
\begin{eqnarray}
(\alpha h\nu)^2=K(h\nu-E_g)\nonumber
\end{eqnarray}
where `K' is the proportionality constant. Fig~(9) shows the variation of 
the band-gap with grain size. The 270 and 480~nm thin 
films have a considerably large band-gap. While one may suspect this to be
due to the existence of ${\rm SnS_2}$ in these samples, it should be noted
that only ${\rm 10\%}$ of the sample is ${\rm SnS_2}$. They are
predominantly SnS films. Also, the linear trends shown in fig~7(a) and (b) 
argue little or no significant contributions are made by ${\rm SnS_2}$. We hence shall assume the large energy band-gaps
to be manifestations of structural and morphological trends of SnS. Before 
commenting on this however, notice the nature of band-gap variation with 
grain size. Experimental data fits to 
\begin{eqnarray}
E_g^{nano}=E_g^{bulk}+{\hbar^2\pi^2 \over 2Mr^2}\label{eeg}
\end{eqnarray}
where `r' is the radius of the nanoparticle and `M' the effective mass of
the system. In other words, the band-gap is inversely proportional to the 
grain size. This fit suggests quantum confinement of the electrons
\cite{eg}. Note that the data points represented by unshaded 
circles in fig~(9) have different `a' lattice parameter. Fig~(10) uses
the same data points of fig~(9), abid in three dimension. We can now see 
the band-gap dependence on both grain size and lattice parameter. 
Mathematically, the band-gap of SnS nano-crystalline thin films is a two 
variable function of grain size and lattice parameter, ${\rm E_g({a,r})}$. 
While the relation between the band-gap and grain size is a well known 
(eqn~\ref{eeg}), there is no known relation between the band-gap and lattice 
parameter (`a'). Hence, more studies would be required to establish the
observed two variable dependency.

\section*{Conclusion}
Tin sulphide films grown on glass substrates at room temperature by thermal
evaporation were found to be nano-crystalline in nature for thicknesses more
than 270~nm. X-Ray diffraction shows that the grain size increased linearly 
with film thickness. Diffraction studies and Electron microscope images show 
that the atoms of SnS are arranged in an layered and oriented manner with 
ortho-rhombic structure and residual tensile stress acting along the 
`a' axis. The films had direct band-gaps ranging from 1.8-2.1~eV
depending on the grain size and lattice parameter. Thus, the study reveals
that the band-gap of SnS is a two variable function, which can be tailored
as per requirement by controlling the grain size and lattice parameter.

\section*{Acknowledgment}
Authors are grateful to the Department of Science and Technology for funding 
this work under research project SR/NM/NS-28/2010.

\newpage
\section*{Figure Captions}
\begin{itemize}
\item[1.] Fig~(a) compares the X-Ray Diffraction Pattern of SnS films of 
various thicknesses. The sample thicknesses and Miller indices are also
indicated. Alongside, (b) shows the variation in lattice parameters with
film thickness. 
\item[2.] TEM micrograph of a 480~nm thin film shows the layered structure of 
SnS thin film. 
\item[3.] The average grain sizes of SnS films were found to be linearly 
proportional to the film thickness.
\item[4.] Raman spectra of SnS thin films of various thicknesses show 
existence of three prominent peaks at ${\rm 170~cm^{-1}}$, ${\rm 238~cm^{-1}}$ 
and ${\rm 330~cm^{-1}}$ (Refer to text). 
\item[5.] XPS peaks of tin and sulphur for 270 and 600~nm thick films are 
visibly different. The 270~nm sample's XPS peaks can be deconvoluted into two 
peaks, indicating presence of SnS (major contribution) and ${\rm SnS_2}$
(minor contributrion). The peaks from ${\rm SnS_2}$ are absent in the 
600~nm sample, thus showing thicker films are of SnS.
\item[6.] Variation of Raman peak positions with film thickness. Both ${\rm
B_{2g}}$ and ${\rm A_g}$ show a decreasing trend with film thickness.
However, ${\rm A_g}$ levels out for film thicknesses above 600~nm (the curve
looks similar to that of fig~1b).
\item[7.] The (a) ${\rm A_g}$ peak position is found to depend on the
lattice parameter `a' while (b) ${\rm B_{2g}}$ peak position shows
dependence on grain size.
\item[8.] Scanning electron micrographs (SEM images) of SnS thin films of 
thicknesses (a) 150, (b) 480, (c) 600 and (d) 900~nm.
\item[9.] Graph shows the variation in band-gap with grain size for as grown 
SnS thin films. The filled circles represent samples that have the same
lattice parameters while unfilled circles represent samples with varying
grain size and lattice parameter. The solid curve is the best fit of 
eqn~(\ref{eeg}) to the data
points. The fit suggests band-gap variation is a result of the electron's 
quantum confinement within the grain.
\item[10.] Three dimension plot shows band gap dependence on lattice
parameter `a' and grain size. The filled and unfilled circles representing
data points are as explained for fig~9.
\end{itemize}
\newpage
\begin{figure}[h!]
\begin{center}
\epsfig{file=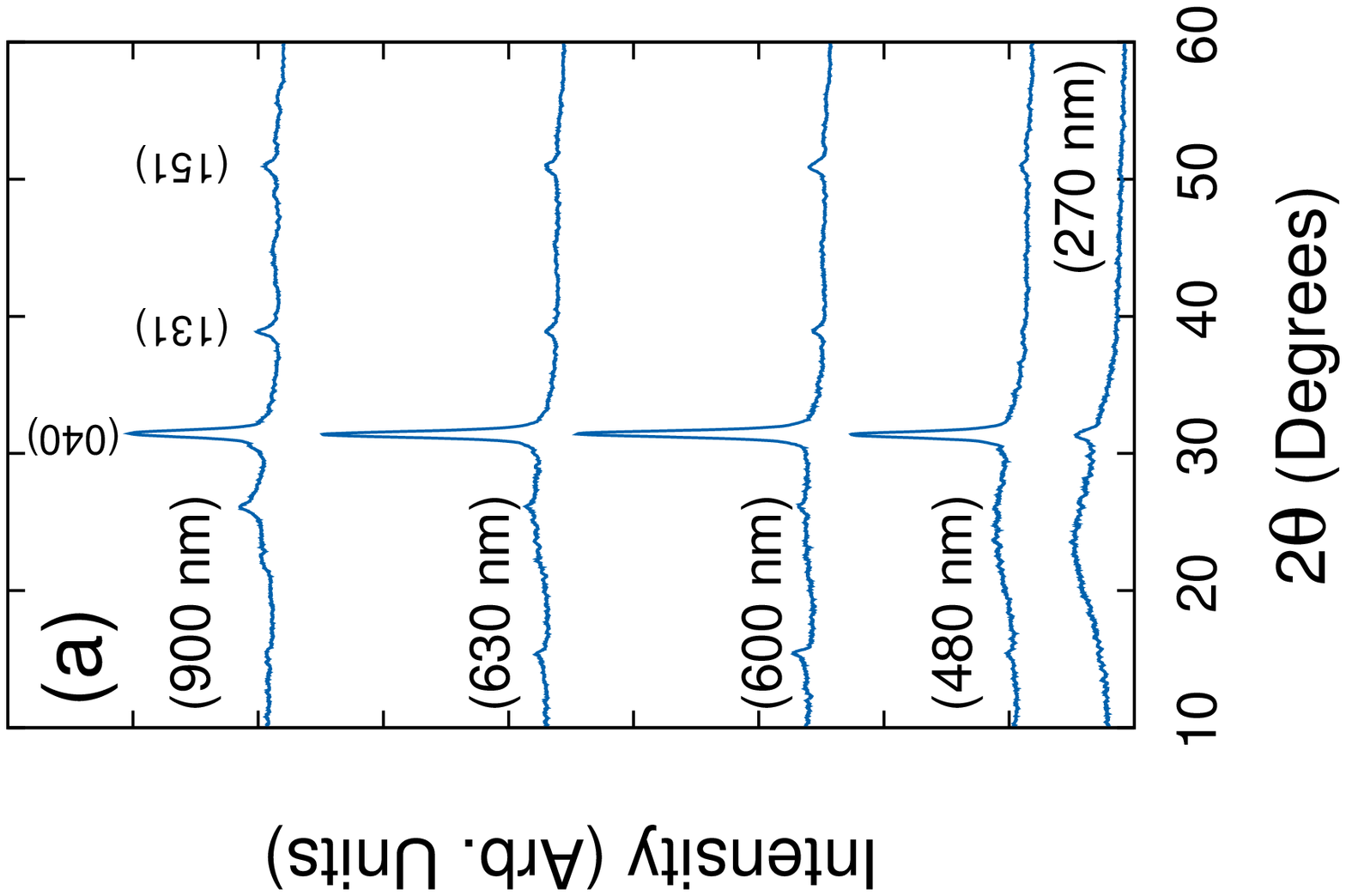, width=3.00in, angle=-90}
\hfil
\epsfig{file=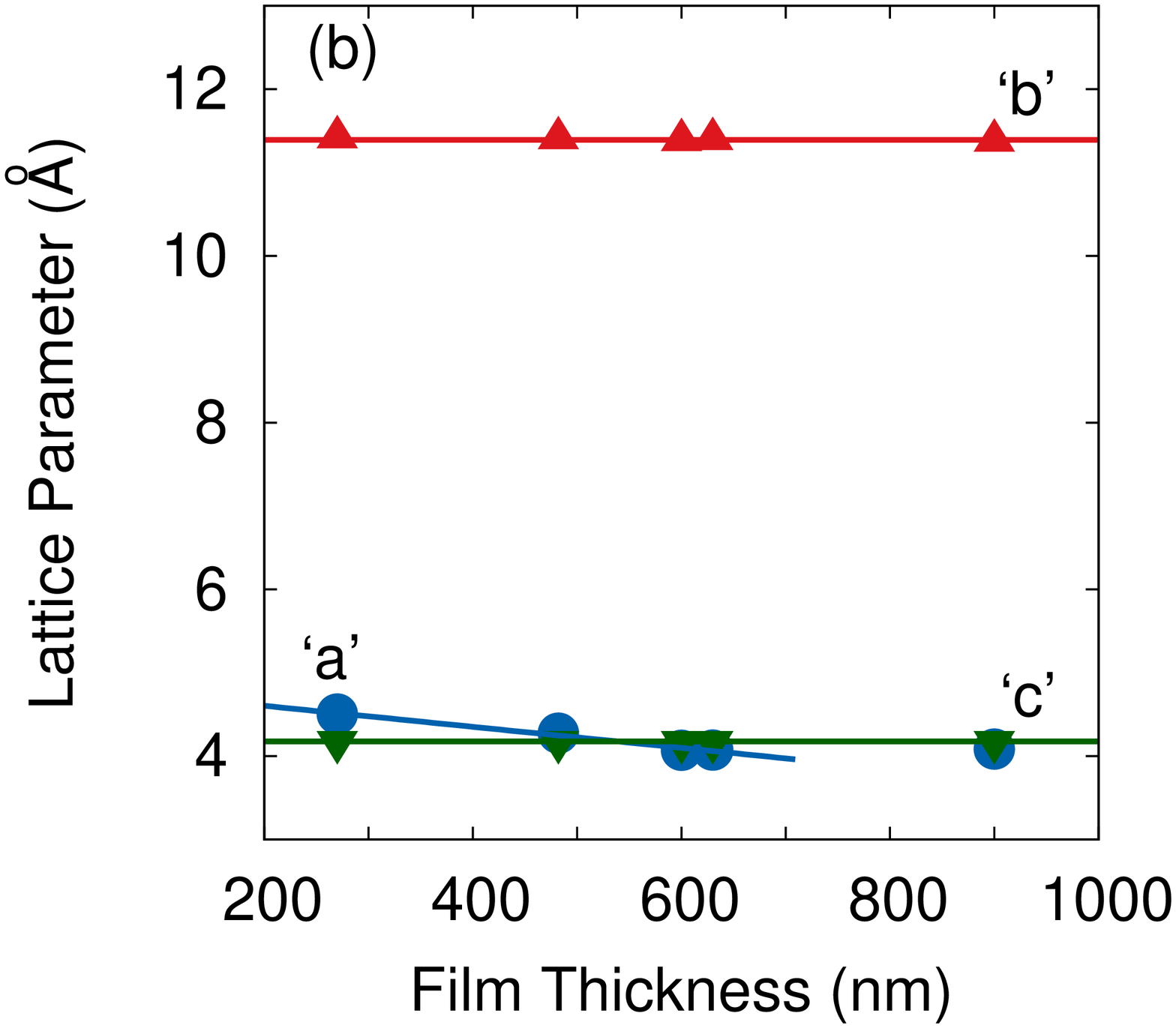, width=3.00in, angle=-90}
\end{center}
\label{fig1}
\caption{\sl Fig~(a) compares the X-Ray Diffraction Pattern of SnS films of 
various thicknesses. The sample thicknesses and Miller indices are also
indicated. Alongside, (b) shows the variation in lattice parameters with
film thickness. Names and peak Miller indices are indicated.}
\end{figure}
\newpage
\begin{figure}[h!]
\begin{center}
\epsfig{file=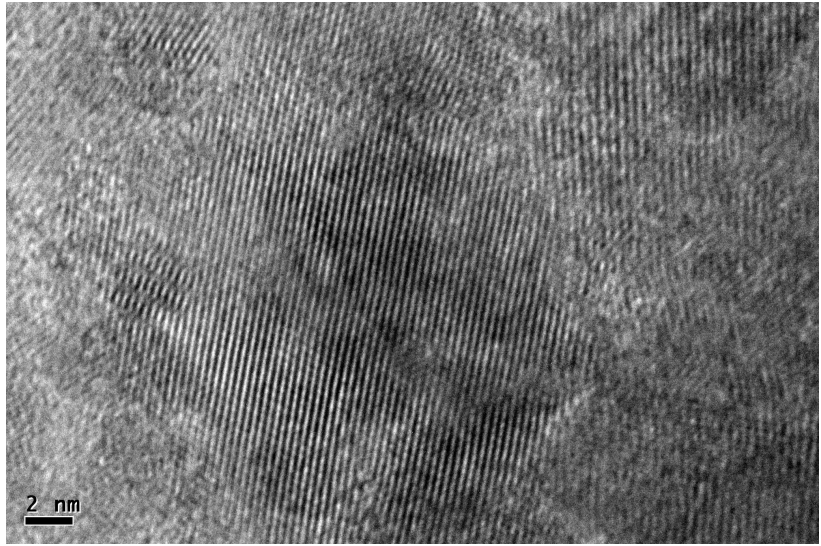, width=4.25in, angle=-0}
\end{center}
\caption{\sl TEM micrograph of a 480~nm thin film shows the layered structure of 
SnS thin film.}
\label{fig2new}
\end{figure}
\newpage
\begin{figure}[h!]
\begin{center}
\epsfig{file=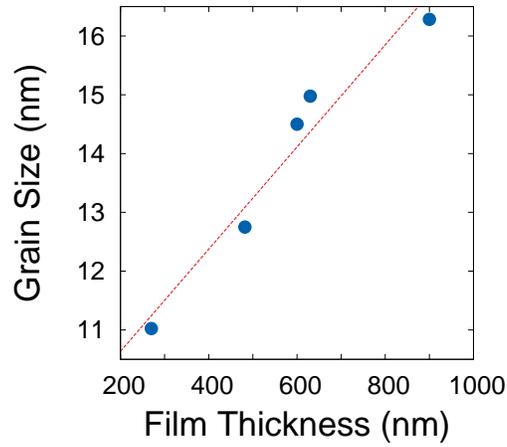, width=2.25in, angle=-90}
\end{center}
\caption{\sl The average grain sizes of SnS films were found to be linearly
proportional to the film thickness.}
\label{fig2}
\end{figure}

\begin{figure}[h!!]
\begin{center}
\epsfig{file=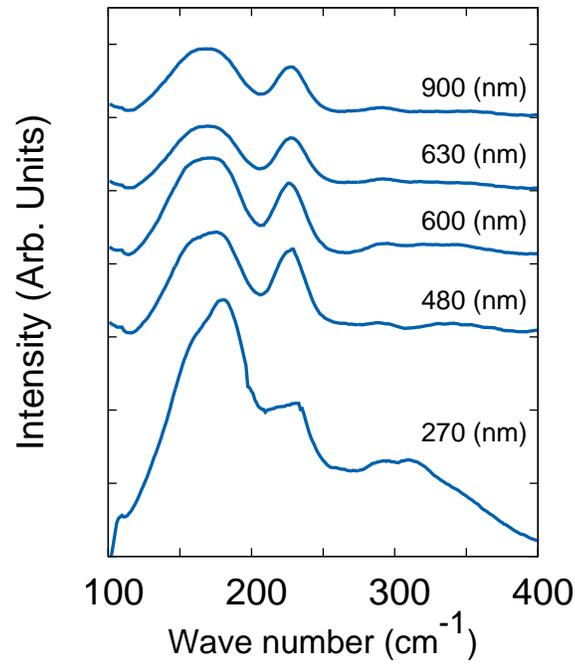, width=3.5in, angle=-90}
\end{center}
\caption{\sl Raman spectra of SnS thin films of various thicknesses show 
existence of three prominent peaks at ${\rm 170~cm^{-1}}$, ${\rm 238~cm^{-1}}$ 
and ${\rm 330~cm^{-1}}$ (Refer to the text). 
}
\label{fig3}
\end{figure}
\begin{figure}[h!]
\begin{center}
\epsfig{file=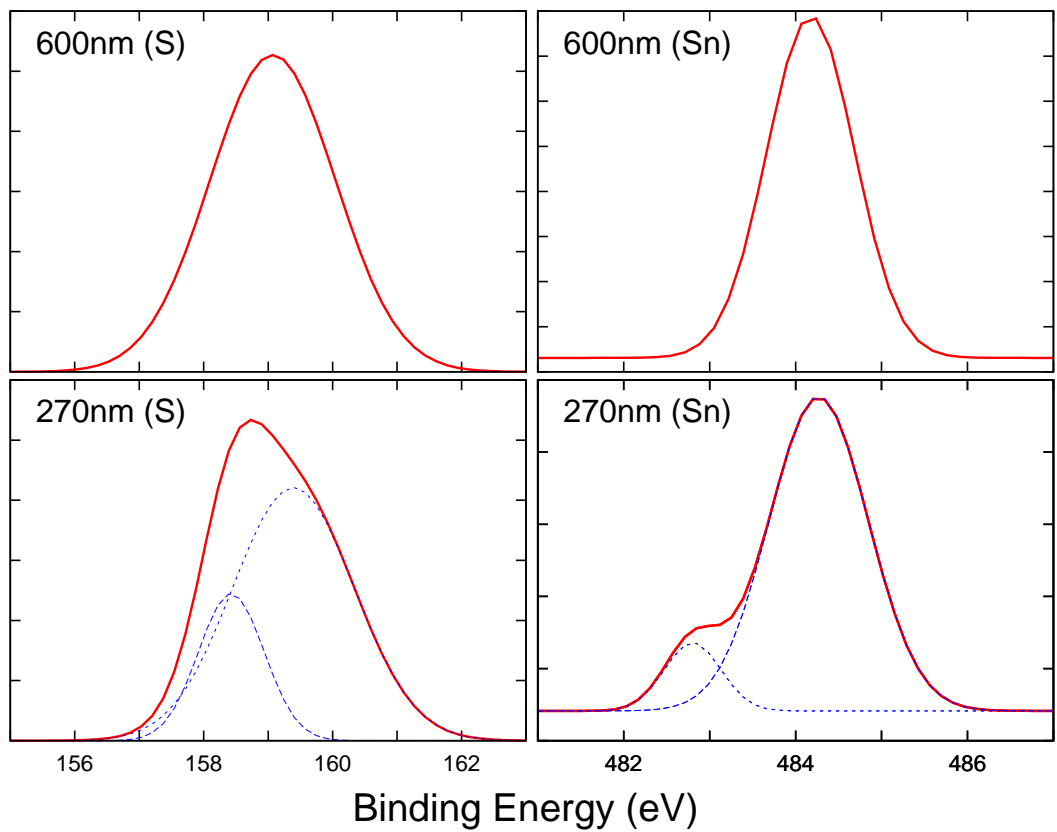, width=4.5in, angle=-90}
\end{center}
\caption{\sl XPS peaks of tin and sulphur for 270 and 600~nm thick films are 
visibly different. The 270~nm sample's XPS peaks can be deconvoluted into two 
peaks, indicating presence of SnS (major contribution) and ${\rm SnS_2}$
(minor contributrion). The peaks from ${\rm SnS_2}$ are absent in the 
600~nm sample, thus showing thicker films are of SnS.}
\label{fig2xps}
\end{figure}

\begin{figure}[h!!]
\begin{center}
\epsfig{file=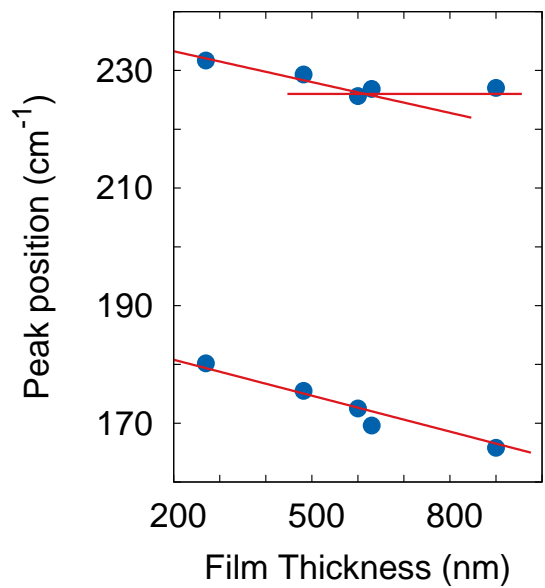, width=3in, angle=-90}
\end{center}
\caption{\sl Variation of Raman peak positions with film thickness. Both ${\rm
B_{2g}}$ and ${\rm A_g}$ show a decreasing trend with film thickness.
However, ${\rm A_g}$ levels out for film thicknesses above 600~nm (the curve
looks similar to that of fig~1b).}
\label{fig4}
\end{figure}
\begin{figure}[h!!]
\begin{center}
\epsfig{file=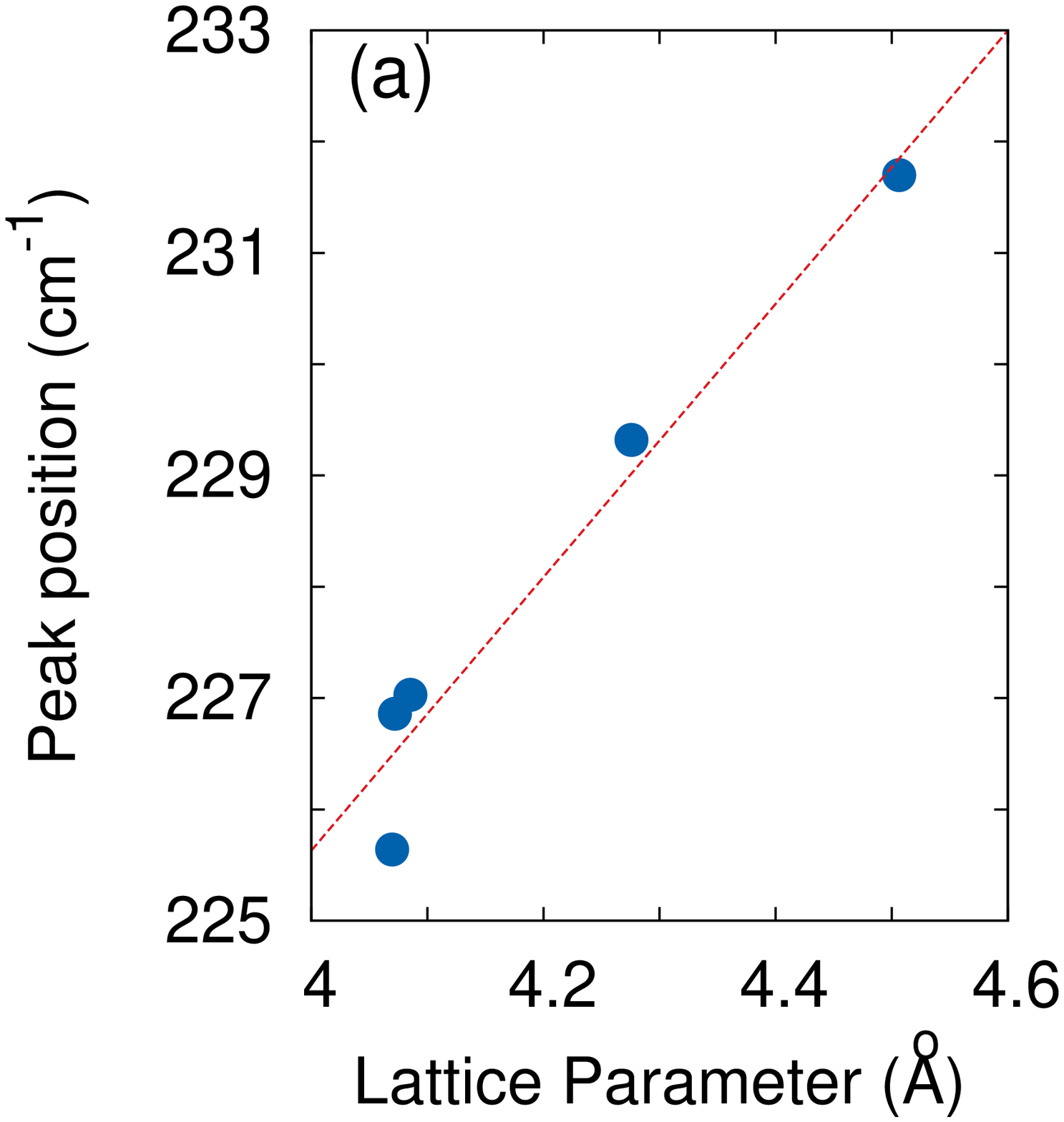, width=3in, angle=-90}
\hfil
\epsfig{file=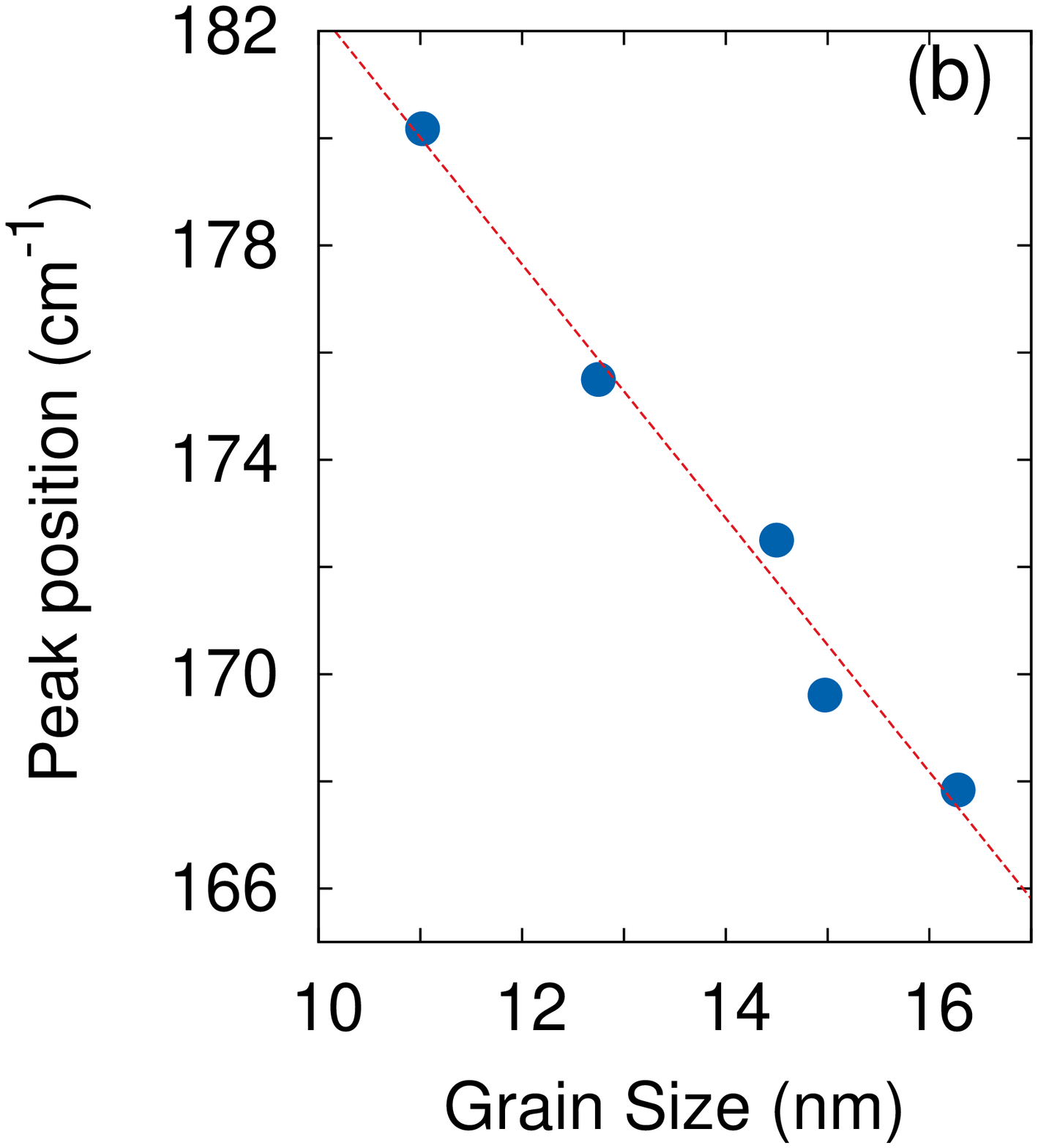, width=3in, angle=-90}
\end{center}
\caption{\sl The (a) ${\rm A_g}$ peak position is found to depend on the
lattice parameter `a' while (b) ${\rm B_{2g}}$ peak position shows
dependence on grain size.
}
\label{fig4}
\end{figure}
\begin{figure}[h!!]
\begin{center}
\epsfig{file=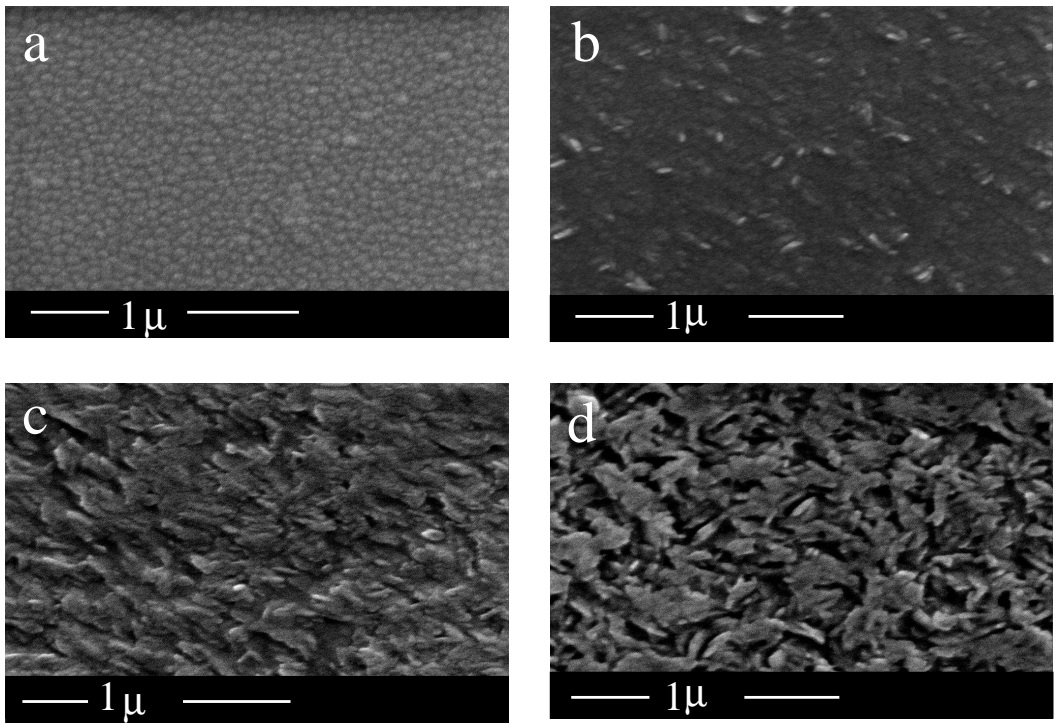, width=5.5in, angle=-0}
\end{center}
\caption{\sl Scanning electron micrographs (SEM images) of SnS thin films of 
thicknesses (a)~150, (b)~480, (c)~600 and (d)~900~nm.
}
\label{fig3sem}
\end{figure}

\begin{figure}[h!!]
\begin{center}
\epsfig{file=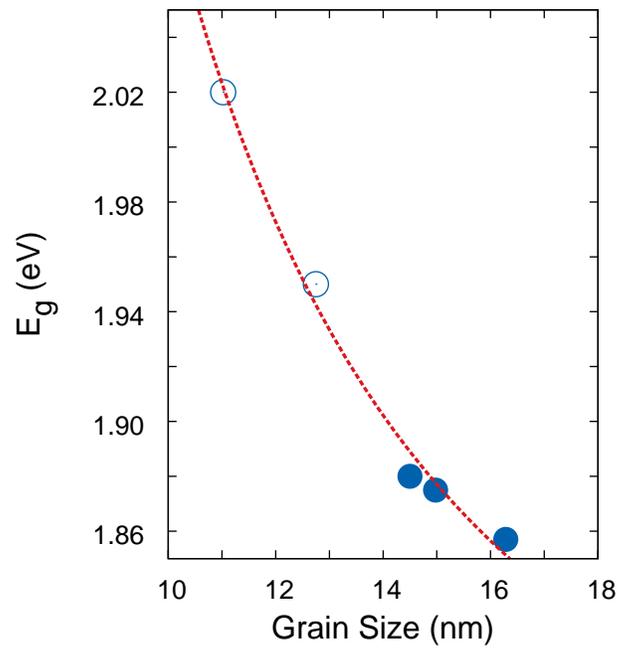, width=3.5in, angle=-90}
\end{center}
\caption{\sl Graph shows the variation in band-gap with grain size for as grown 
SnS thin films. The filled circles represent samples that have the same
lattice parameters while unfilled circles represent samples with varying
grain size and lattice parameter. The solid curve is the best fit of 
eqn~(\ref{eeg}) to the data
points. The fit suggests band-gap variation is a result of the electron's 
quantum confinement within the grain.
}
\end{figure}

\begin{figure}[h!!]
\begin{center}
\epsfig{file=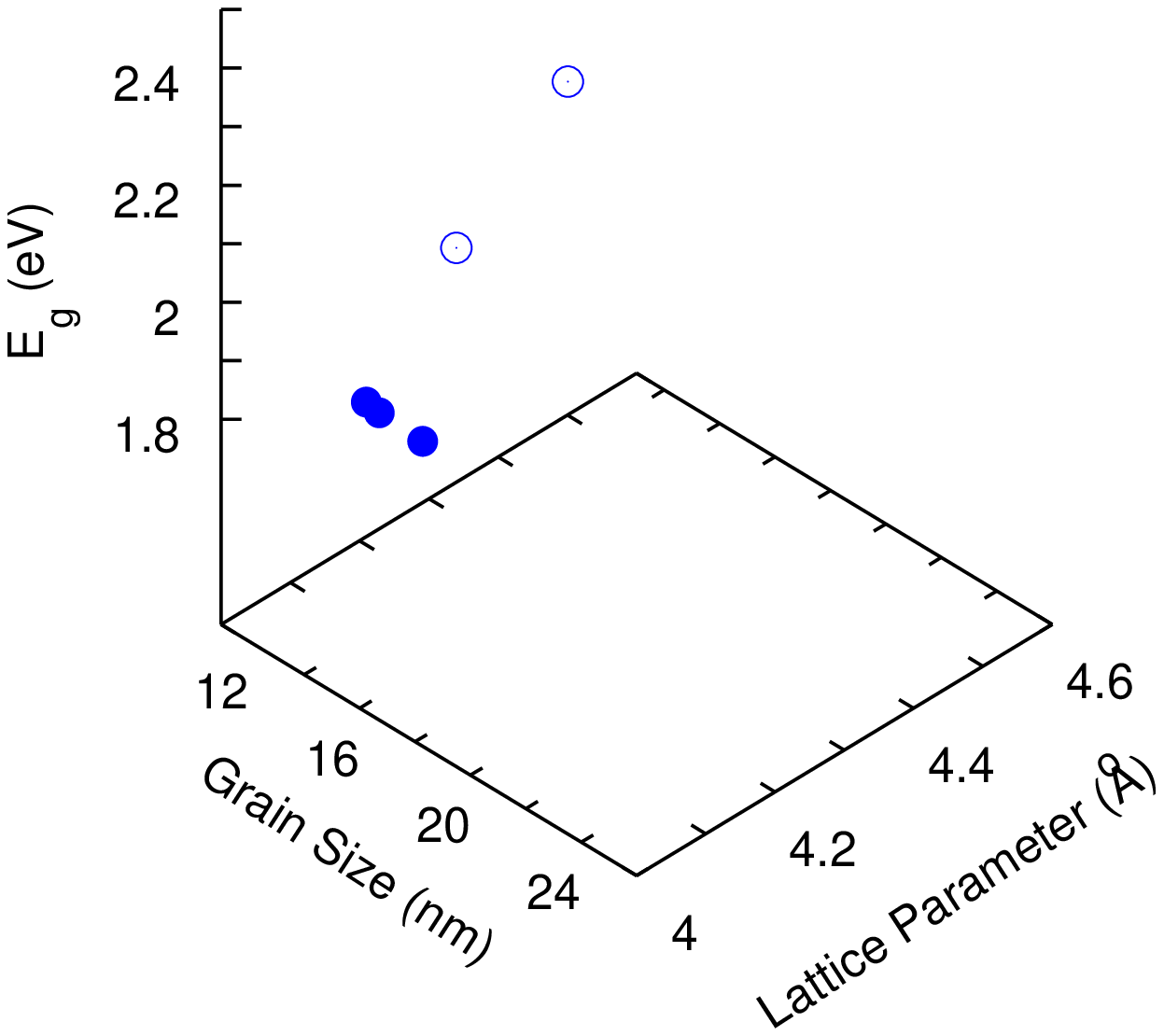, width=5in, angle=-90}
\end{center}
\caption{\sl Three dimension plot shows band gap dependence on lattice
parameter `a' and grain size. The filled and unfilled circles representing
data points are as explained for fig~7.}
\end{figure}


\begin{thebibliography}{99}
\bibitem{amartya} Chakrabarti A., Lu J., McNamara A.M., Kuta L.M., Stanley S.M., Xiao Z., Maguire J.A., Hosmane N.S. 
Tin(IV) sulfide:Novel nanocrystalline morphologies. Inorg Chim Acta, 2011, 374:627
\bibitem{radot} Yue G.H., Wang L.S., Wang X., Chen Y.Z., Peng D.L. Characterization and Optical Properties of the Single Crystalline SnS Nanowire Arrays
Nanoscale Res Lett, 2009, 4:359 
\bibitem{patil} Patil S.G., Tredgold R.H. Electrical and photoconductive properties of ${\rm SnS_2}$ crystals.
 J Pure Appl Phys, 1971, 4:718
\bibitem{zhi} Wang Z., Qu S., Zeng X., Liu J., Zhang C., Tan F., Jin L., Wang Z.
The application of SnS nanoparticles to bulk heterojunction solar cells. J Alloys Comp, 2009, 482: 203
\bibitem{noguchi} Noguchi H., Setiyadi A., Tanamora H., Nagatomo T.,  Omoto O. Characterization of vacuum-evaporated tin sulfide film for solar cell materials.
 Sol. Energy Mater. Sol. Cells, 1994, 35:325
\bibitem{nikol} Nikolic P.M., Miljkovic Lj., Mihajlovic P., Lavrencic B.
Splitting and coupling of lattice modes in the layer compound SnS. J Phys C: Solid State Phys, 1977, 10: L289
\bibitem{albanesi} Makinistian L., Albanesi E.A. Study of the hydrostatic pressure on the orthorhombic IV-VI compounds including many-body effects.
 J Comput Mater Sci, 2011, 50:2872
\bibitem{crc} ``CRC Handbook of Chemistry and Physics (1993-1994)'' ${\rm
74^{th}}$ Edn., Ed. Linde D.R. (Boca RA Raton, FL: CRC Press).
\bibitem{nahass} El-Nahass M.M., Zeyada N.M., Aziz M.S., El-Ghamaz N.A. Optical properties of thermally evaporated SnS thin films.
Opt Mater, 2002, 20:159
\bibitem{hartman} Hartman K., Johnson J.L., Bertoni M.I., Recht D., 
Aziz M.J., Scarpulla M.A., Buonassis T. SnS thin-films by RF sputtering at room temperature. Thin Solid
Films, 2011, 519:7421
\bibitem{nair} Nair M.T.S., Nair P.K. Simplified chemical deposition technique for good quality SnS thin films.
Semicond Sci Technol, 1991, 6:132
\bibitem{ghaza} Ghazali A., Zainal Z., Hussein M.Z., Kassim A. Cathodic electrodeposition of SnS in the presence of EDTA in aqueous media.
Sol. Energy Mater. Sol. Cells, 1998, 55:237
\bibitem{thanga} Thangaraju B., Kaliannan P. Spray Pyrolytic deposition and characterization of SnS and ${\rm SnS_2}$.
 J Phys D: Appl Phys, 2000, 33:1054
\bibitem{gao} Gao C., Shen H., Sun L. Preparation and properties of zinc blende and orthorhombic SnS films by chemical bath deposition.
 Appl Surf Sci, 2011, 257:6750
\bibitem{sohila} Sohila S., Rajalakshmi M., Ghosh C., Arora A.K., Muthamizhchelvan C. Optical and Raman scaterring studies on SnS nanoparticles.
J Alloy Compd, 2011, 509:5843
\bibitem{yue} Yue G.H., Wang W., Wang L.S., Wang X., Yan P.X., Chen Y., Peng D.L.
The effect of annealing temperature on physical properties of SnS films. J Alloy Compd, 2009, 474:445 
\bibitem {Feng} Jiang F., Shen H., Gao C., Liu B., Lin L., Shen Z. Preparattion and properties of SnS film grown by two-stage process.
Appl Surf Sci, 2011, 257:4901
\bibitem {Cullity} Cullity B.D., Stock S.R., ``Elements of X-Ray Diffraction",
 ${\rm 3^{rd}}$ Ed., Prentice-Hall Inc (NJ, 2001)
\bibitem {Lucovsky} Lucovsky G., Mikkelsen J.C., Liang W.Y., White R.M. Optical phonon anisotropies in the layer
 crystals ${\rm SnS_2}$ and ${\rm SnSe_2}$. Phys Rev B, 1976, 14:1664
\bibitem{cross} Band I.M., Kharitonov Y.I., Trzhaskovskaya M.B. Photoionization cross sections and photoelectron angular distributions
 for x-ray line energies in the range 0.132–4.509 keV targets: 1 ≤ Z ≤ 100. At Data Nucl. Data Tables, 1979, 23:443
\bibitem{sigben} Briggs D. ``Handbook of X-ray and Ultra-voilet Photo-electron Spectroscopy", Perkin-Elmer Corportation.
 Physical Electronics Division, 1978
\bibitem {Chandrasekhar} Chandrasekhar H.R., Humphreys R.G., Zwick U.,Cardona M.
 Infrared and Raman spectra of the IV-VI compounds SnS and SnSe. Phys Rev B, 1977, 15:2177  
\bibitem {hyunchul} Choi H.C., Jung Y.M., Kim S.B. Size effects in Raman spectra of ${\rm TiO_2}$ nanoparticles.
J Vib Spectrosc, 2005, 37:33
\bibitem{s8} Rajalakshmi M., Arora A.K., Bendre B.S., Mahamuni S. Optical phonon confinement in zinc oxide nanoparticles.
 J Appl Phys, 2000, 87:2445
\bibitem{s9} Yang C.L., Wang J.N., Ge W.K., Guo L., Yang S.H., Shen D.Z. Enhanced Ultraviolet emission and optical properties in polyvinyl
pyrrolidone surface modified ZnO quantum dots. J Appl Phys, 2001, 90:4489
\bibitem{s10} Guo L., Yang S., Yang C., Yu P., Wang J., Ge W., Wong G.K.L. Highly monodisperse polymer-capped ZnO:nanoparticles: Preparation
and optical properties. Appl Phys Lett, 2000, 76:2901
\bibitem{s11} Alim K.A., Fonoberov V.A., Balandin A.A. Origin of the optical phonon frequency shifts in ZnO quantum dots.
 Appl Phys Lett, 2005, 86:053103-1
\bibitem {kuninori} Kitahara K., Ishii T., Suzuki J., Bessyo T., Watanabe N. Characterization of Defects and Stress in
 Polycrystalline Silicon Thin Films on Glass Substrates by Raman Microscopy. INT J Spectrosc, 2011, 2011:1
\bibitem{zhu2} Zhu J.S., Lu X.M., Jiang W., Tian W., Zhu M., Zhang M.S., Chen X.B., Liu X., Wang Y.N. 
Optical study on the size effects in BaTiO3 thin films. J Appl Phys, 1997, 81:1392 
\bibitem {Ehm} Ehm L., Knorr K., Dera P., Krimmel A., Bouvier P., Mezouar M. Pressure induced structural phase transition in the
IV-VI semiconductor SnS. J Phys:Condens Mater, 2004, 16:3545
\bibitem{r3} Devika M., Reddy N.K., Ramesh K., Ganesan V., Gopal E.S.R. ,Reddy K.T.R. Influence of substrate temperature on surface
 structure and electrical resistivity of the evaporated tin sulphide films. Appl Surf Sci, 2006, 253:1673
\bibitem{beer} Streetman B. ``Solid State Electronic Devices", ${\rm
4^{th}}$ Ed., PHI (New Delhi, 1995).
\bibitem {Devika} Devika M., Reddy N.K., Prashantha M., Ramesh K., Reddy S., Hahn Y.B., Gunasekhar K.R. The physical properties
of SnS films grown on lattice-matched and amorphous substrates. Phys Status Solidi A, 2010, 207:1864 
\bibitem{tauc} Tauc J. Absorption edge and internal electric fields in amorphous semiconductors.
 Mat. Res. Bull., 1970, 5:721
\bibitem{mott} Mott N.F., Davis E.A. ``Electronic Process in
Non-Crystalline Materials", Clarendon Press (Oxford, 1979).
\bibitem{pank} Pankove J.I. ``Optical Processes in Semiconductors", PHI, NY
(1971).
\bibitem{tauc2} Tauc J. ``Amorphous and Liquid Semiconductors", J. Tauc
Ed., Plenum, London (1974). 
\bibitem{clark} Clark A.H. ``Polycrystalline and Amorphous Thin Films and
Devices", Kazmerski L. Ed., Academic Press (NY, 1980).
\bibitem{eg} Brus L.E. Electron–electron and electron‐hole interactions in small semiconductor crystallites: The size dependence
of the lowest excited electronic state. J Chem Phys, 1984, 80:4403

\end {thebibliography}
\end{document}